\documentclass[slac_one]{revtex4}
\usepackage{graphicx}
\usepackage{fancyhdr}
\usepackage{color}
\begin{document}

\title{  
{\small 2005 International Linear Collider Workshop - Stanford U.S.A.}\\ 
\vspace{12pt}
The impact of beamstrahlung on precision measurements at CLIC}
\author{Asesh K.~Datta}
\affiliation{MCTP, University of Michigan, Ann Arbor, MI 48109-1040, USA}
\author{Kyoungchul Kong\footnote{This talk was given by K.~Kong, 
describing past and ongoing work performed in collaboration with 
the other authors.}, Konstantin T.~Matchev}
\affiliation{Physics Department, University of Florida, Gainesville, FL 32611, USA}
\begin{abstract}
We review two commonly used approximate treatments of beamstrahlung at lepton colliders. 
We discuss the applicability of those approximations and show that they are bound to fail at
very high energy $e^+e^-$ colliders such as CLIC. We model the beamstrahlung effects at CLIC 
by an exact numerical solution to the evolution equation for beamstrahlung and compare to the two
approximations. We also discuss the impact of beamstrahlung on precision measurements of new 
physics parameters. As an illustration we consider Kaluza-Klein lepton production in
Universal Extra Dimensions and study the deterioration of the endpoint lepton energy 
spectrum caused by beamstrahlung.
\end{abstract}

\maketitle

\thispagestyle{fancy}

\section{INTRODUCTION} 
\label{sec:intro}

The fixed center of mass energy $E_{CM}$ is a celebrated virtue of lepton colliders. 
In contrast to hadron colliders, where the parton-level $E_{CM}$ varies from 
event to event, the additional kinematic constraint allows for a number of 
new measurements. Some well-known examples are: 
the presence of sharp endpoints in the distributions of various kinematic observables,
the possibility of threshold scans, measurements of the missing mass, 
the angular distributions at production, etc.
The availability of all these new experimental tools allows for precise 
determination of new physics parameters and is the primary motivation for pursuing
the construction of a next generation lepton collider.

One should keep in mind, however, that in reality a fraction of the beam energy 
is always lost due to the effects of beamstrahlung and bremsstrahlung, so that the
true center of mass energy in the event is often displaced from its nominal value.
The purpose of the present note is to investigate to what extent these effects 
can represent a problem for precision measurements of new physics parameters 
at very high energy $e^+e^-$ colliders such as CLIC.
To this end, in Sec.~\ref{sec:approx} we first review the two commonly used 
parametrizations of beamstrahlung. Then in Sec.~\ref{sec:applicability}
we discuss the applicability of those approximations for lepton colliders with
sub-TeV $E_{CM}$  (such as ILC) as well as $E_{CM}$ in the multi-TeV range 
(such as CLIC). We find that the latter case falls outside the range of validity 
of the two approximations, and an exact treatment of beamstrahlung effects 
is required. We solve numerically the evolution equation for beamstrahlung
using the exact expression for the Sokolov-Ternov transition probability
(see Sec.~\ref{sec:approx}). In Sec.~\ref{sec:muons} we consider the 
example of Kaluza-Klein muon production in Universal Extra Dimensions (UED) and 
use our numerical results from Sec.~\ref{sec:applicability} to investigate
the deterioration of the endpoint lepton energy spectrum caused by beamstrahlung.
The same analysis can be readily applied to smuon production in supersymmetry
as well. 

\section{ANALYTICAL APPROACHES TO BEAMSTRAHLUNG}
\label{sec:approx}

\begin{figure*}[t]
\begin{picture}(0,0)%
\includegraphics{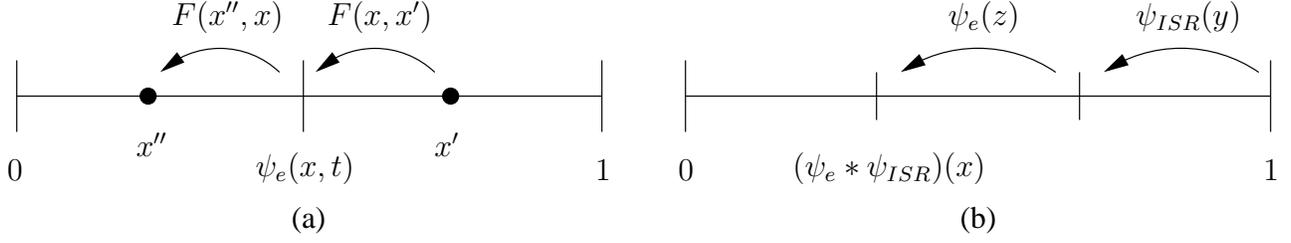}%
\end{picture}%
\setlength{\unitlength}{3947sp}%
\begingroup\makeatletter\ifx\SetFigFont\undefined%
\gdef\SetFigFont#1#2#3#4#5{%
  \reset@font\fontsize{#1}{#2pt}%
  \fontfamily{#3}\fontseries{#4}\fontshape{#5}%
  \selectfont}%
\fi\endgroup%
\begin{picture}(7947,1489)(241,-644)
\put(1051,-136){\makebox(0,0)[lb]{\smash{\SetFigFont{12}{14.4}{\rmdefault}{\mddefault}{\updefault}{\color[rgb]{0,0,0}$x''$}%
}}}
\put(1276,689){\makebox(0,0)[lb]{\smash{\SetFigFont{12}{14.4}{\rmdefault}{\mddefault}{\updefault}{\color[rgb]{0,0,0}$F(x'',x)$}%
}}}
\put(2926,-136){\makebox(0,0)[lb]{\smash{\SetFigFont{12}{14.4}{\rmdefault}{\mddefault}{\updefault}{\color[rgb]{0,0,0}$x'$}%
}}}
\put(3936,-286){\makebox(0,0)[lb]{\smash{\SetFigFont{12}{14.4}{\rmdefault}{\mddefault}{\updefault}{\color[rgb]{0,0,0}$1$}%
}}}
\put(2251,689){\makebox(0,0)[lb]{\smash{\SetFigFont{12}{14.4}{\rmdefault}{\mddefault}{\updefault}{\color[rgb]{0,0,0}$F(x,x')$}%
}}}
\put(6151,689){\makebox(0,0)[lb]{\smash{\SetFigFont{12}{14.4}{\rmdefault}{\mddefault}{\updefault}{\color[rgb]{0,0,0}$\psi_e(z)$}%
}}}
\put(7331,689){\makebox(0,0)[lb]{\smash{\SetFigFont{12}{14.4}{\rmdefault}{\mddefault}{\updefault}{\color[rgb]{0,0,0}$\psi_{ISR}(y)$}%
}}}
\put(1801,-261){\makebox(0,0)[lb]{\smash{\SetFigFont{12}{14.4}{\rmdefault}{\mddefault}{\updefault}{\color[rgb]{0,0,0}$\psi_e (x,t)$}%
}}}
\put(5176,-261){\makebox(0,0)[lb]{\smash{\SetFigFont{12}{14.4}{\rmdefault}{\mddefault}{\updefault}{\color[rgb]{0,0,0}$(\psi_e * \psi_{ISR})(x)$}%
}}}
\put(241,-286){\makebox(0,0)[lb]{\smash{\SetFigFont{12}{14.4}{\rmdefault}{\mddefault}{\updefault}{\color[rgb]{0,0,0}$0$}%
}}}
\put(4456,-286){\makebox(0,0)[lb]{\smash{\SetFigFont{12}{14.4}{\rmdefault}{\mddefault}{\updefault}{\color[rgb]{0,0,0}$0$}%
}}}
\put(8131,-286){\makebox(0,0)[lb]{\smash{\SetFigFont{12}{14.4}{\rmdefault}{\mddefault}{\updefault}{\color[rgb]{0,0,0}$1$}%
}}}
\end{picture}
\caption{(a) $F(x,x')$ corresponds to the source and $F(x'', x)$ to the sink for the evolution 
of the spectral function $\psi_e(x,t)$.
(b) The final energy of an electron results from a convolution of beamstrahlung and bremsstrahlung 
(see Sec.~\ref{sec:convolution}).}
\label{fig:beamst}
\end{figure*}

For physics purposes we need to know the energy distribution $\psi_e(x,t)$ 
of the colliding electrons, where $x \equiv E/E_0$ is the electron energy 
as a fraction of the nominal beam energy $E_0$. Neglecting electrons from 
pair-production, the normalization condition on $\psi_e$ is 
\begin{equation}
\int_0^1 dx\, \psi_e(x,t) = 1 .
\label{norm}
\end{equation} 
Assuming that the emission of a photon takes place on an infinitesimally short
time scale, the interference between successive radiation processes 
is negligible and the evolution of the electron spectral function $\psi_e$ 
is described by the following rate equation (see Figure~\ref{fig:beamst}a):
\begin{equation}
\frac{\partial\psi_e (x,t)}{\partial t} = - \int_0^x d x'' F(x'',x)\psi_e(x,t) 
                                    + \int_x^1 d x' F(x,x')\psi_e(x',t)\ .
\label{rate_eqn}
\end{equation}
Here $F(x,x')$ is the spectral function of radiation, i.e. 
$F(x,x')$ is the transition probability (per unit time) for 
an electron at $x'$ to move into the energy interval $(x,x+dx)$. 
Obviously, $F(x,x')=0$ if $x\geq x'$. Pulling out $\psi_e(x,t)$,
which is independent of $x''$, outside the $x''$ integral,
leaves the integral quantity 
\begin{equation}
\nu(x) = \int_0^x dx'' F(x'',x)\ ,
\label{nux}
\end{equation}
which represents the average number of photons radiated per 
unit time by the electron with an instantaneous energy $x$.
The transition probability $F(x,x')$ derived by Sokolov and 
Ternov~\cite{Sokolov:1986nk} is given by
\begin{eqnarray}
F(x,x')     &=& \frac{\nu_{cl}\kappa}{x x'} f(\xi, \eta)\,, \label{STfunction} \\
f(\xi,\eta) &=& \frac{3}{5\pi}\frac{1}{1+\xi\eta}\left [ \int_\eta^\infty du K_{5/3}(u) + 
                                      \frac{\xi^2\eta^2}{1+\xi\eta} K_{2/3}(\eta)  \right ]\,,
\label{fksieta}
\end{eqnarray}
where $\xi\equiv3x'\Upsilon/2$, $\eta=\kappa\left( 1/x - 1/x' \right)$ and the constant 
$\kappa\equiv 2/(3\Upsilon)$ is introduced for convenience. 
$K_\nu$ is the modified Bessel function and $\nu_{cl}$ 
is the number of photons per unit time calculated by the 
classical theory of radiation. By definition, 
this is also the limit of $\nu(x)$ for $x\rightarrow 0$:
\begin{equation}
\nu_{cl} \equiv \nu(x=0) = \frac{5}{2\sqrt{3}}\frac{\alpha^2}{r_e \gamma_0}\Upsilon\,,
\end{equation}
where $r_e$ is the classical electron radius, $\alpha$ is the fine structure constant, 
$\gamma_0=E_0/(m_ec^2)$ and the dimensionless Lorentz invariant parameter 
\begin{equation}
\Upsilon=\frac{5}{6} \frac{r_e^2 \gamma_0 N}{\alpha \sigma_z (\sigma_x +\sigma_y)}
\end{equation}
has a value specific to any given collider, as it depends on the 
total number of particles $N$ in a gaussian bunch and 
the rms sizes $\sigma_x$, $\sigma_y$, $\sigma_z$ of the gaussian beam. 

\subsection{Yokoya-Chen (YC) approximation}
\label{sec:YC}

It is clear that the Sokolov-Ternov (ST) spectral function (\ref{STfunction}) is too complicated
to allow a general analytic solution to Eq.~(\ref{rate_eqn}). Therefore, one has to resort to 
some kind of an approximation. In 1989 Yokoya and Chen~\cite{Yokoya:1989jb,Chen:1991wd} 
suggested the following simple approximation to (\ref{fksieta}):
\begin{equation}
f(\xi, \eta) \approx \frac{1}{\Gamma(1/3)} \eta^{-2/3}e^{-\eta}\, ,
\label{fYC}
\end{equation}
with the additional assumption
\begin{equation}
\nu(x) = \int_0^x d x'' F(x, x'') \approx \nu_{cl} = \, constant .
\label{nucl}
\end{equation}
One can then solve the rate equation (\ref{rate_eqn}) by Laplace transformation
and obtain an analytical solution:
\begin{equation}
\psi(x,t) = e^{-\nu_{cl}t} \left [ \delta(1-x) + 
                \frac{e^{-\eta_x}}{1-x} h(\eta_x^{1/3}\nu_{cl}t) \right ]\, ,
\label{YCsol}
\end{equation}
where $\eta_x \equiv \kappa (1/x - 1)$ and 
\begin{equation}
h(u) = \frac{1}{2\pi i} \int_{\lambda-i\infty}^{\lambda+i\infty} exp(up^{-1/3}+p)dp 
     = \sum_{n=1}^{\infty}\frac{u^n}{n!\Gamma(n/3)}
\end{equation}
with $\lambda > 0$ and $0 \leq u \leq \infty$. The first term in Eq.~(\ref{YCsol}) 
represents the electron population unaffected by radiation. The $n$-th term in 
the Taylor expansion of the second term in (\ref{YCsol}) corresponds to the process 
of $n$-photon emission. The final electron energy distribution $\psi_e(x) $ 
due to beamstrahlung is obtained by suitable time averaging of 
the time-dependent solution (\ref{YCsol}) to account for the duration of bunch overlap, 
for example:
\begin{equation}
\psi_e(x) \equiv \frac{2}{l}\int_{0}^{l/2}dt\ \psi_e(x,t) 
        = \frac{1}{N_{cl}} \left [ (1-e^{-N_{cl}}) \delta(1-x) 
                     + \frac{e^{-\eta_x}}{1-x} \bar{h}(x) \right ]\,,
\label{psiYC}
\end{equation}
where $l=2\sqrt{3}\sigma_z$ is the effective length of the incoming bunch and 
$N_{cl}=\nu_{cl}l/2$ is the average number of photons radiated 
per particle during the entire collision of the $e^+e^-$ beams. 
The function $\bar{h}(x)$ in the second term is
\begin{equation}
\bar{h}(x) = \sum_{n=0}^{\infty}\frac{\eta_x^{n/3}}{n!\Gamma(n/3)}\gamma(n+1,N_{cl})\,,
\end{equation}
where $\gamma(n,x)$ is the incomplete Gamma function
\begin{equation}
\gamma(n,x) = \int_0^x dy\, e^{-y} y^{n-1}\,.
\end{equation}
The resulting solution (\ref{psiYC}) is in good agreement with ABEL simulation data at least
up to $\Upsilon \sim 0.44$ \cite{Chen:1991wd}.

\subsection{Consistent Yokoya-Chen (CYC) approximation}
\label{sec:CYC}

Later on, M.~Peskin pointed out \cite{Peskin:1999pk} that the two assumptions 
(\ref{fYC}) and (\ref{nucl}) behind the YC approximation
are incompatible with the normalization condition (\ref{norm}) 
of the energy spectral function. To rectify the problem, Peskin proposed 
the so-called ``consistent Yokoya-Chen'' approximation as an alternative to (\ref{fYC}):
\begin{equation}
\tilde{f}(\xi,\eta) = \frac{x'}{x}\frac{1}{\Gamma(1/3)} \eta^{-2/3}e^{-\eta}\, .
\label{fCYC}
\end{equation}
The rate equation (\ref{rate_eqn}) can then be solved in exactly the same way and the 
resulting analytic solution is
\begin{equation}
\psi_e(x) = e^{-N_\gamma} \left ( \delta(x-1) + \frac{e^{-\eta_x}}{x(1-x)} h(N_\gamma \eta^{1/3}_x) \right )\,, 
\label{psiCYC}
\end{equation}
where $N_\gamma = \sqrt{3}\sigma_z\nu_{cl}(1+\Upsilon^{2/3})^{-1/2}$. 
This solution not only preserves the probability sum rule (\ref{norm})
but also agrees with Guinea Pig simulation for $\Upsilon = 0.104$ (NLC500) and 
$\Upsilon = 0.299$ (NLC1000) and numerically is not very different from 
the YC solution (\ref{psiYC})  for $\Upsilon \ll 1$ \cite{Peskin:1999pk}.
However, for large $\Upsilon$, the two solutions are significantly different,
see Fig.~\ref{fig:psi} and Sec.~\ref{sec:applicability} below.

\begin{figure*}[t]
\centering
\includegraphics[width=82mm]{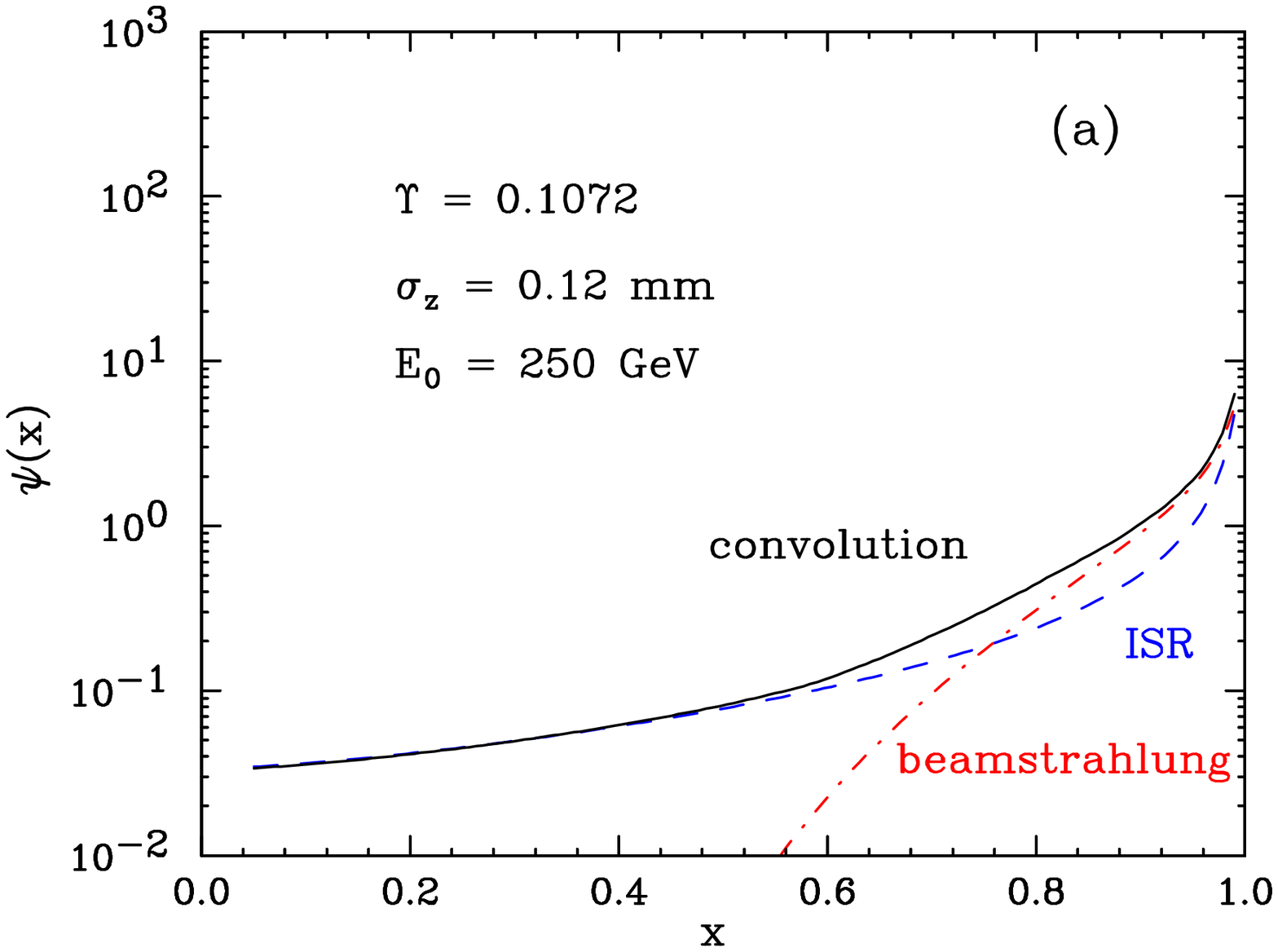}~~
\includegraphics[width=82mm]{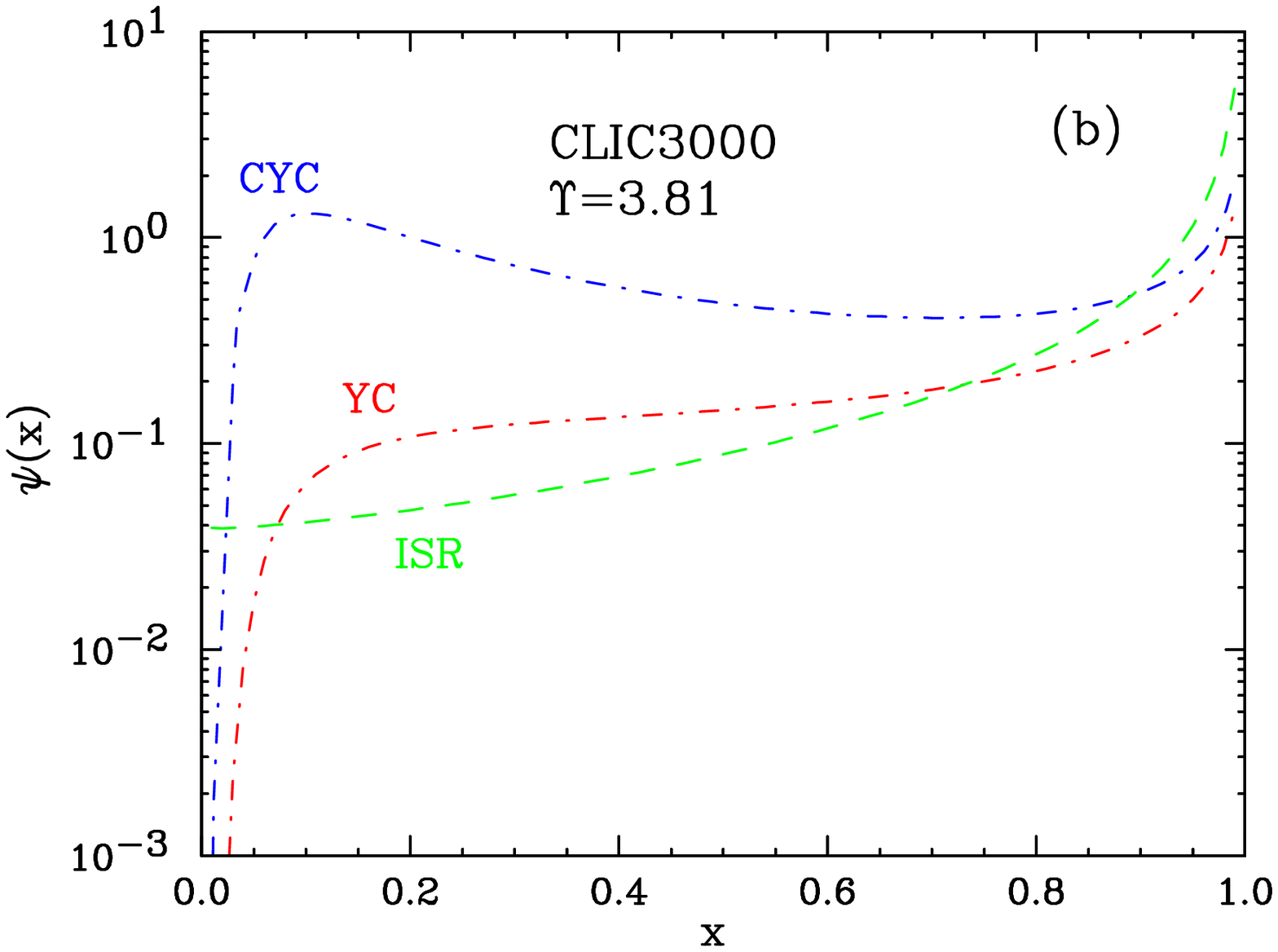}
\caption{
(a) Electron energy distributions due to beamstrahlung (dot-dashed, red),
ISR (dashed, blue) and their convolution (solid), for an ILC with the parameters as shown.
The YC and CYC approximations (\ref{psiYC}) and (\ref{psiCYC}) are almost identical
in this case. (b) Electron energy distributions due to beamstrahlung (dot-dashed)
and ISR (dashed, green), for CLIC with the shown parameters.
The YC (CYC) solution is shown in red (blue).}
\label{fig:psi}
\end{figure*}

\subsection{Convolution with bremsstrahlung}
\label{sec:convolution}

Another source of electron energy losses is bremsstrahlung, or initial state radiation (ISR), 
parametrized with its own energy distribution function $\psi_{ISR}$ \cite{Kuraev:1985hb}.
The probability to end up with an electron of energy $x$ 
is a convolution of ISR and beamstrahlung (see Figs.~\ref{fig:beamst}b and \ref{fig:psi}a):
\begin{equation}
\psi_{ISR}*\psi_{e} = \psi_{e}*\psi_{ISR} 
                 = \int_0^1 dy dz\, \psi_{ISR} (y) \psi_e (z) \delta(x - y z) 
                 = \int_{x}^1 \frac{dz}{z}\, \psi_{ISR}\left ( \frac{x}{z} \right ) \psi_{e}(z)\ .
\end{equation}

\section{VALIDITY RANGE OF THE APPROXIMATE SOLUTIONS}
\label{sec:applicability}

\begin{figure*}[t]
\centering
\includegraphics[width=82mm]{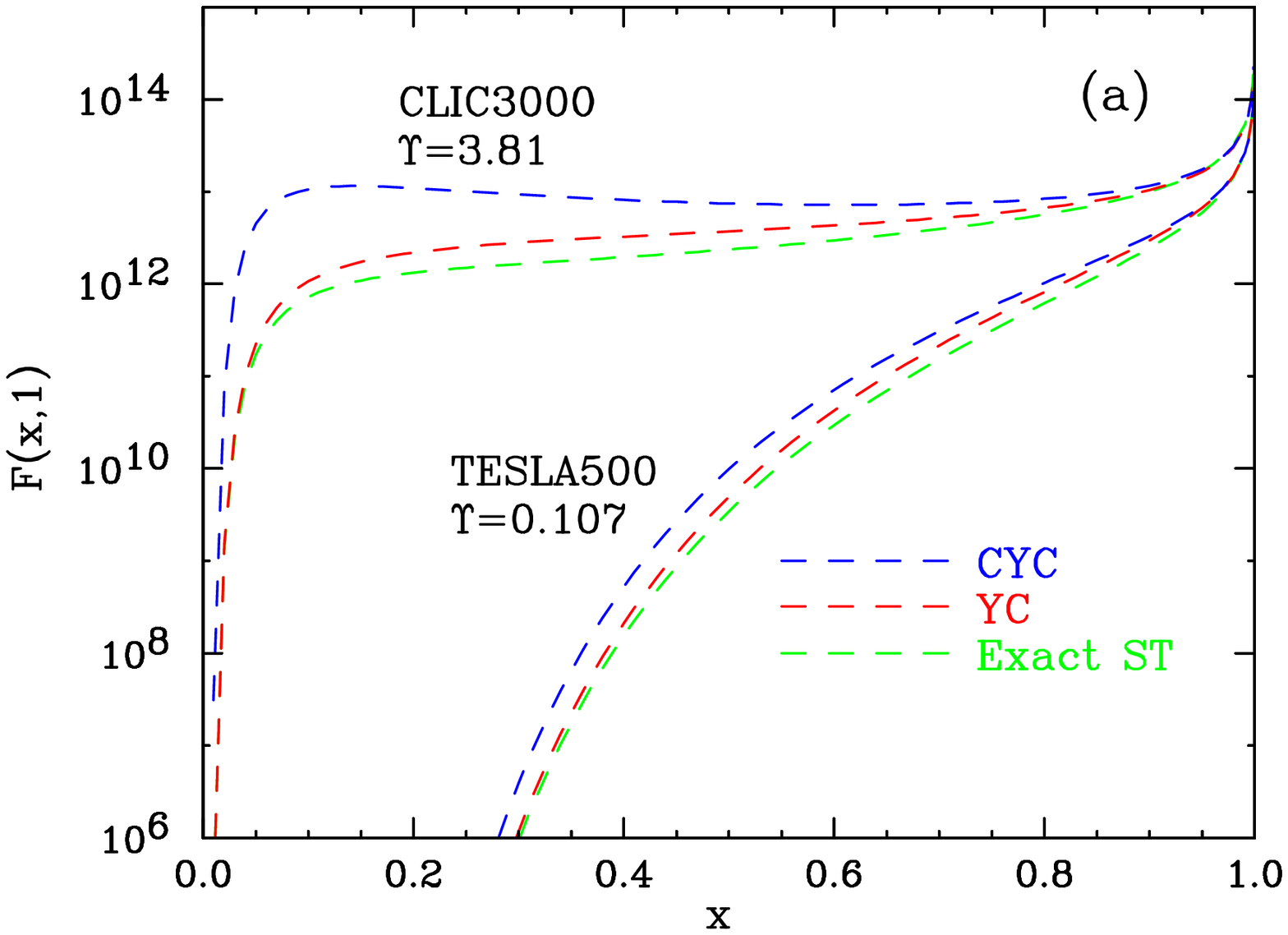}
\includegraphics[width=82mm]{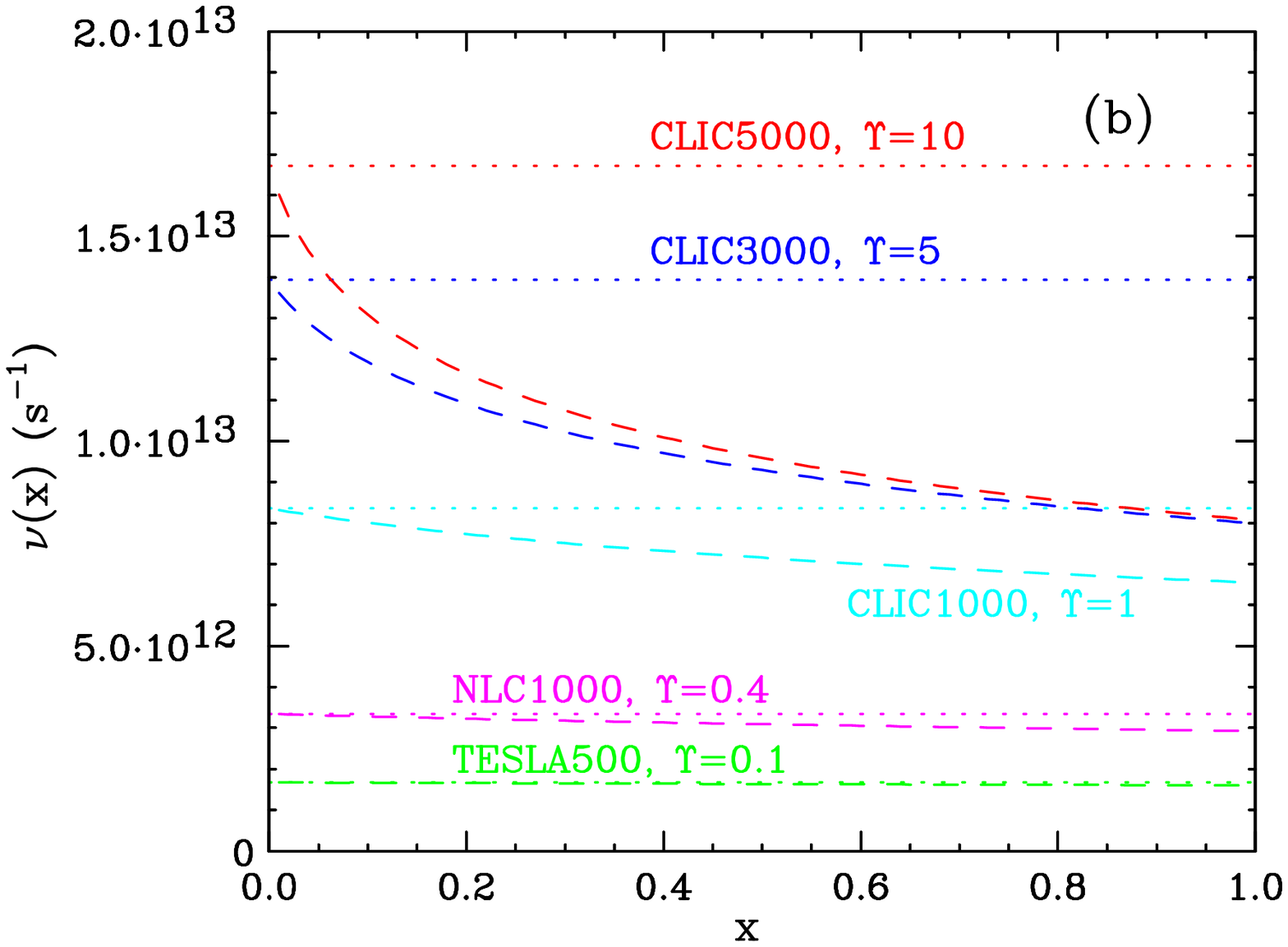}
\caption{(a) The transition probability $F(x,1)$ as a function of $x$,
using the exact expression (\ref{fksieta}) or the approximations
(\ref{fYC}) and (\ref{fCYC}), for two different values of $\Upsilon$.
(b) The function $\nu(x)$ plotted for various values of $\Upsilon$.}
\label{fig:approx}
\end{figure*}

In order to test the validity of the YC and CYC approximations, in
Fig.~\ref{fig:approx}a we plot the transition probability $F(x,1)$ 
as a function of $x$,
using the exact expression (\ref{fksieta}) or the approximations
(\ref{fYC}) and (\ref{fCYC}), for two different values of $\Upsilon$.
First of all, we notice that as $\Upsilon$ gets large, one enters the
quantum regime of radiation, where emitting hard photons becomes more 
and more likely. The large tail at low $x$ appearing in the case of CLIC
could become an annoying spoiler for physics studies --- 
an issue we will look at in more detail in the next Section.
We also see that fixing the normalization problem of
the YC approximation comes at a price: the improved CYC approximation is
numerically further away from the exact ST spectral function (\ref{STfunction})
and the discrepancy becomes more pronounced at large $\Upsilon$.
Conversely, while the YC approximation does a better job in approximating 
(\ref{STfunction}), it not only suffers from the normalization problem, but 
also relies on the approximation $\nu(x)\approx \nu_{cl}=const$, which fails at large
$\Upsilon$, as evidenced in Fig.~\ref{fig:approx}b.
There we plotted $\nu(x)$ and the corresponding $\nu_{cl}$ 
for different values of $\Upsilon$. We see that for small
$\Upsilon$ the $\nu(x)$ function (\ref{nux}) is pretty constant
and the YC approximation is valid, but for the larger $\Upsilon$ values 
which are relevant at CLIC, $\nu(x)$ is far from being a constant, and the
YC approximation will fail.

Fig.~\ref{fig:approx} reveals that in order to get a reliable parametrization
of beamstrahlung for large $\Upsilon$ (i.e.~at very high energy 
$e^+e^-$ colliders such as CLIC), we need to improve on the 
two existing approximations. We tried, but were unable to find
a better approximation to (\ref{fksieta}), which would still allow
for an analytic solution in closed form. We therefore resorted to
a numerical solution of the original rate equation (\ref{rate_eqn})
with the full ST formula (\ref{STfunction}).
This solution was then used for the physics study 
presented in Sec.~\ref{sec:muons}.

\section{MUON ENERGY SPECTRUM FROM KK-MUON PRODUCTION}
\label{sec:muons}

In order to study the effect of beamstrahlung on physical observables, 
we consider the production of level 1 Kaluza-Klein muons $\mu_1^\pm$
in UED~\cite{Appelquist:2000nn,Cheng:2002ab}, which then decay to the 
lightest KK particle (LKP), in this case the KK partner $\gamma_1$ of the 
photon. The signature is two opposite sign muons and missing energy, 
which is analogous to smuon production in supersymmetry models
with stable neutralino LSP. At hadron colliders, the two scenarios
can be confused~\cite{Cheng:2002ab,Battaglia:2005zf,Smillie:2005ar}, 
but lepton colliders allow for an easy and straightforward 
discrimination~\cite{Battaglia:2005zf}.

\begin{figure*}[t]
\centering
\includegraphics[width=82mm]{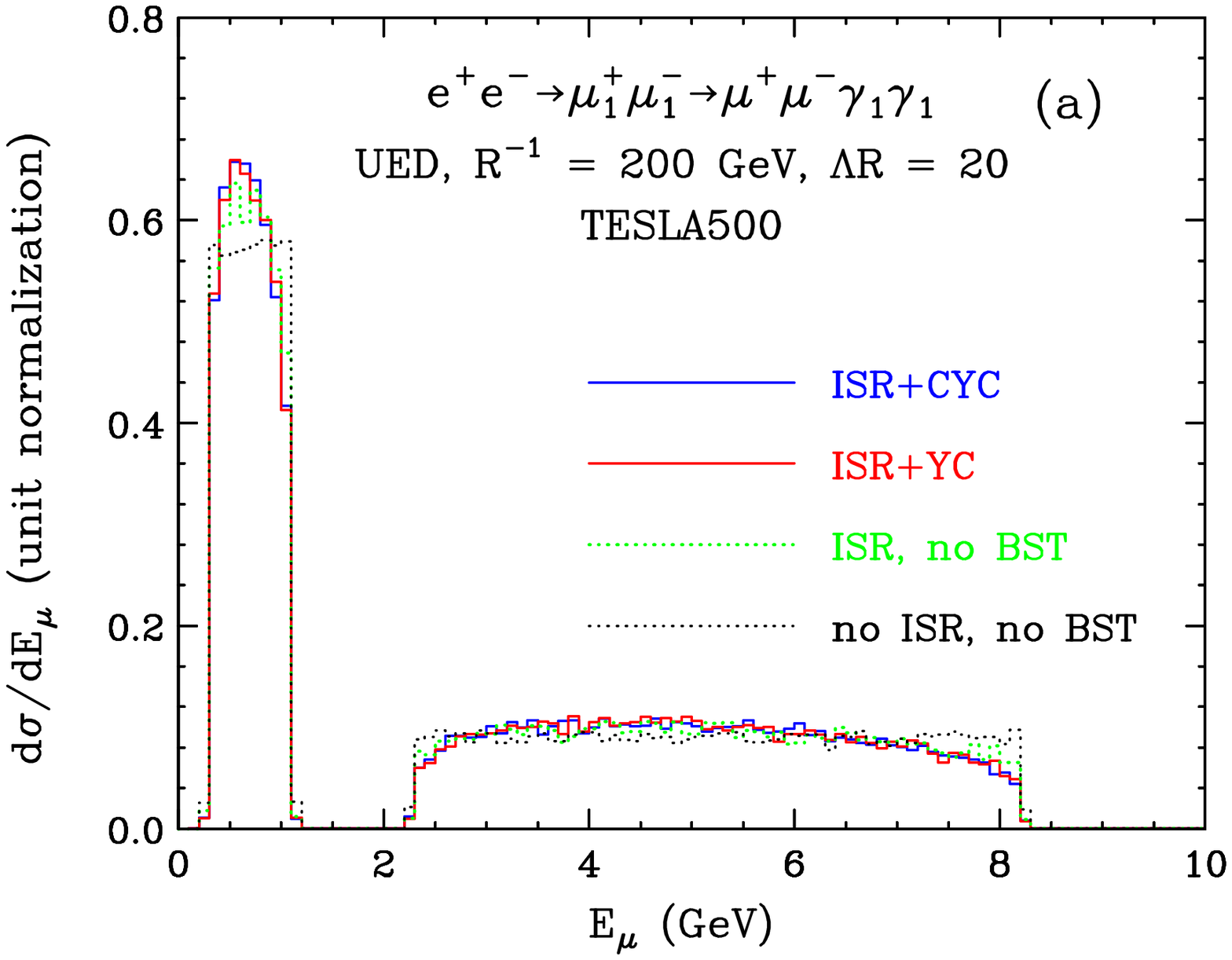}
\includegraphics[width=82mm]{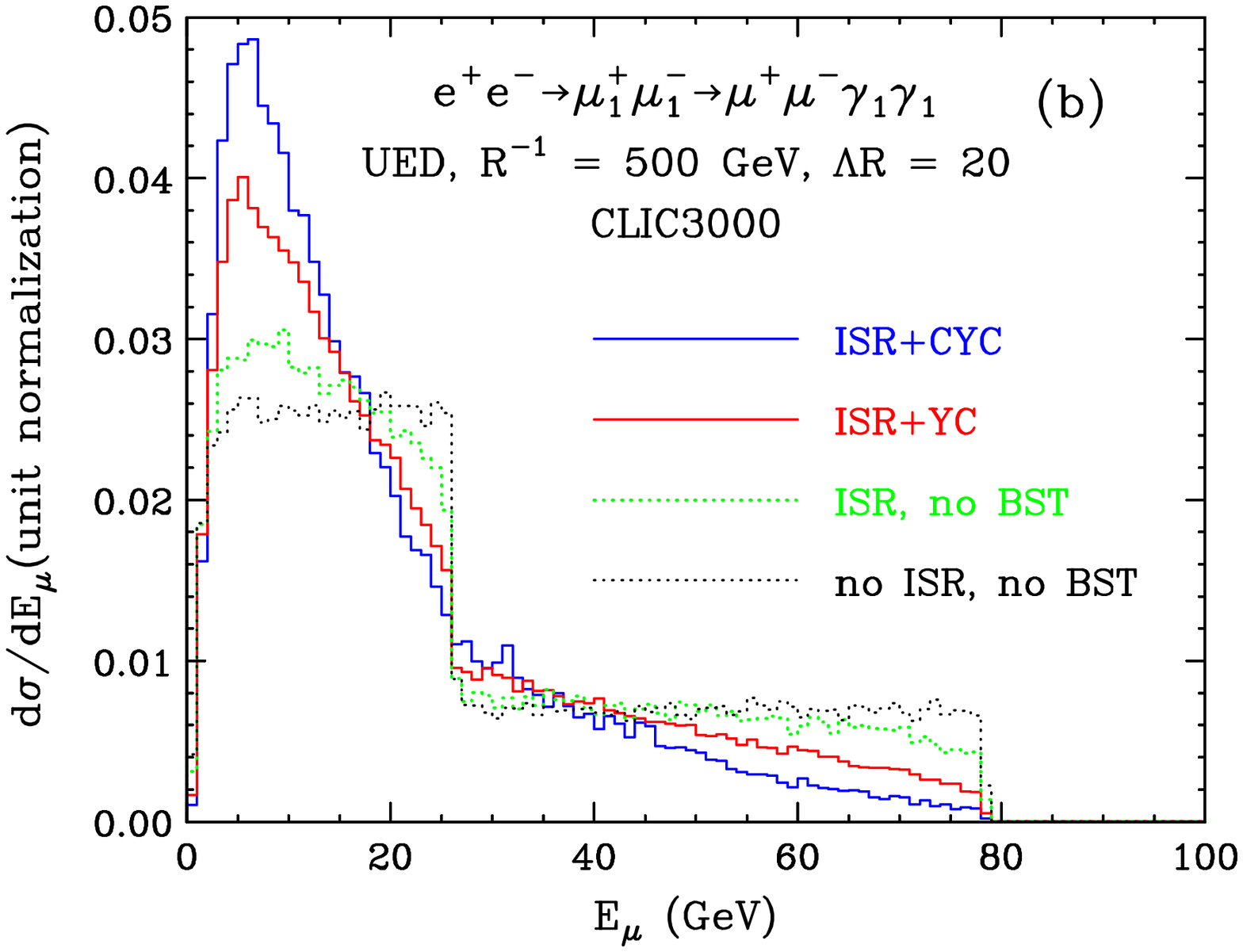}
\caption{(a) Muon energy spectrum from KK muon production in UED at an 
$e^+e^-$ collider with $E_{CM}=500$ GeV and $\Upsilon = 0.107$.
The left (right) box-like distribution is 
due to SU(2)-singlet (SU(2)-doublet) $\mu_1^\pm$ production.
The dotted lines ignore the beamstrahlung, with the effects from 
ISR shown in green. The red (blue) lines include beamstrahlung 
according to the YC (CYC) approximation.
(b) The same as (a), but for $R^{-1}=500$ GeV at CLIC with $E_{CM}=3000$ GeV
and $\Upsilon = 3.81$.}
\label{fig:muons}
\end{figure*}

Two important observables are the lower, $E_{min}$, 
and upper, $E_{max}$, endpoints of the muon energy spectrum 
(see Fig.~\ref{fig:muons}). They can be used to extract the masses 
of the particles involved in the decay as follows:
\begin{equation}
E_{max/min} = \frac{1}{2} M_{\mu_1} 
      \left ( 1- \frac{M^2_{\gamma_1}}{M^2_{\mu_1}} \right ) \gamma (1 \pm \beta)\, ,
\end{equation}
where $M_{\mu_1}$ and $M_{\gamma_1}$ are the KK muon and LKP masses, respectively,
$\gamma=1/(1-\beta^2)^{1/2}$, and $\beta$ is the $\mu_1$ boost.
Fig.~\ref{fig:muons}a confirms that at low $E_{CM}$ the effects from
beamstrahlung are rather small, and furthermore, the YC and CYC approximations 
are in agreement. However, Fig.~\ref{fig:muons}b reveals that at higher energy 
colliders, beamstrahlung tends to wash out the endpoint, the degree of deterioration 
being quite sensitive to the approximation used.

\begin{figure*}[t]
\centering
\includegraphics[width=82mm]{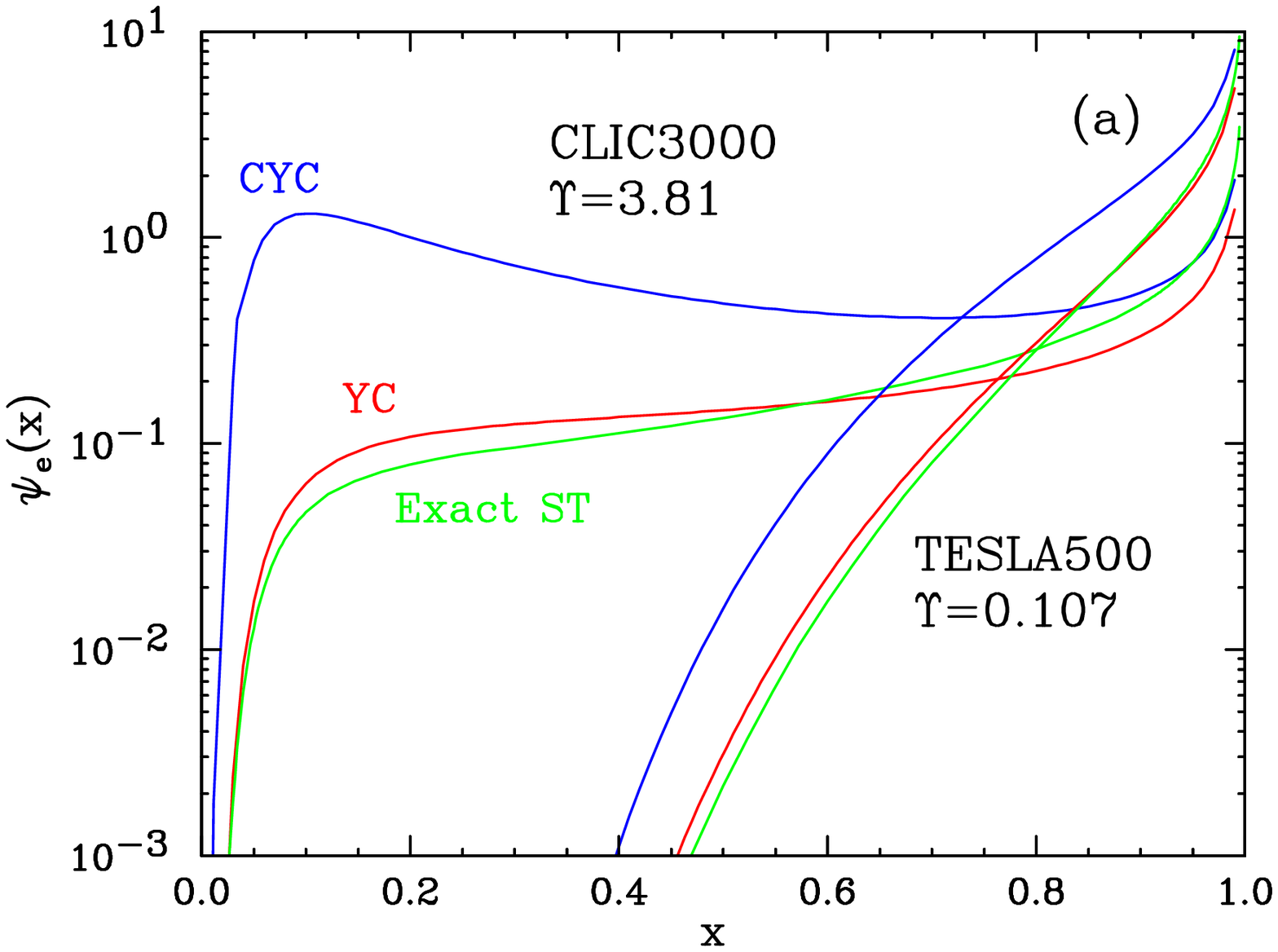}
\includegraphics[width=80mm]{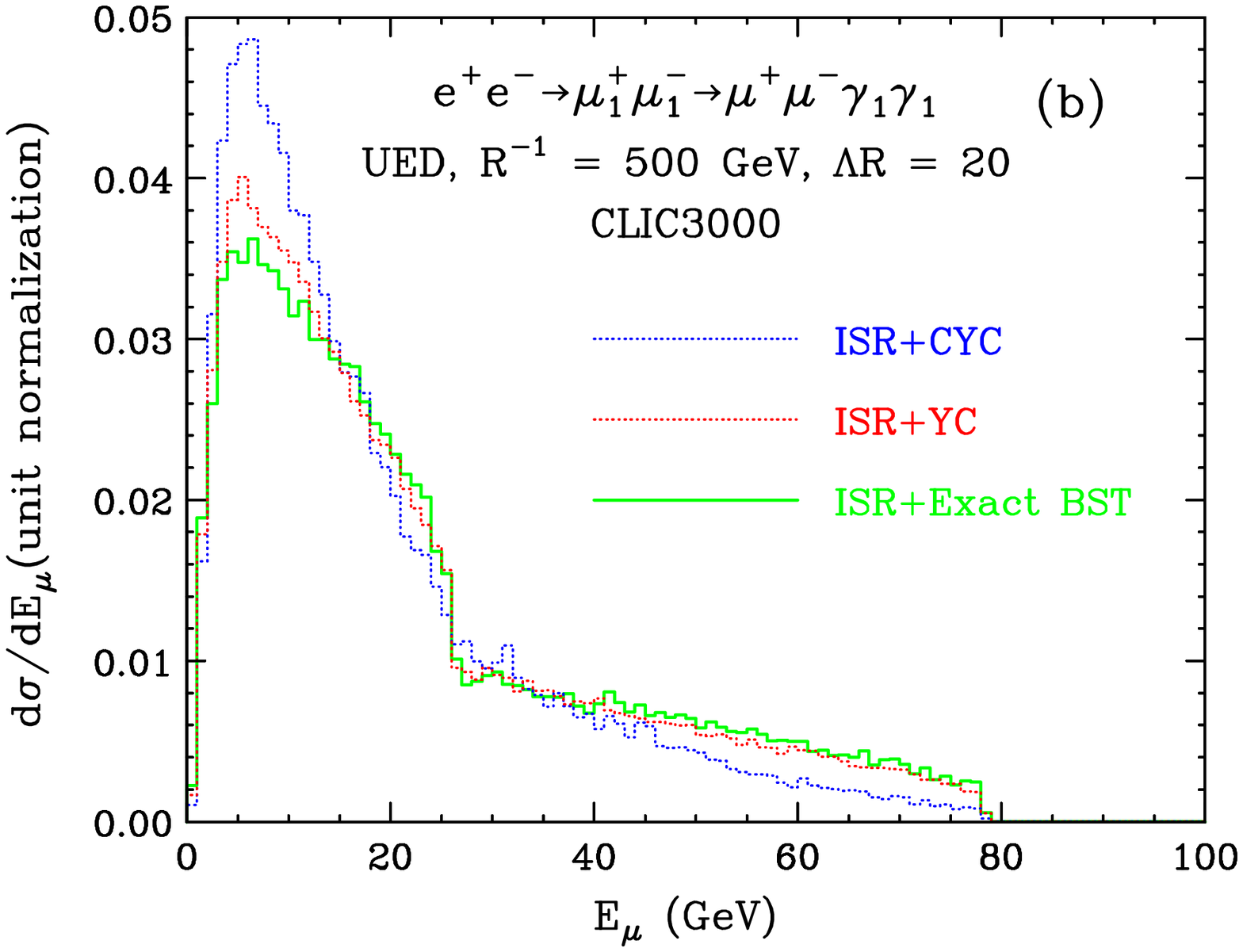}
\caption{(a) Electron energy distributions resulting from the numerical solution (green)
or the YC (red) and CYC (blue) approximations, for $\Upsilon = 3.81$ and $\Upsilon = 0.107$. 
(b) The same as Fig.~\ref{fig:muons}b, but including the exact solution for beamstrahlung.}
\label{fig:numsol}
\end{figure*}

Since we already expect neither approximations to be reliable at large $E_{CM}$
(where $\Upsilon$ is also large), we solved Eq.~(\ref{rate_eqn}) numerically.
The result, together with the YC and CYC approximations, is shown in Fig.~\ref{fig:numsol}a,
for $\Upsilon=0.107$ (TESLA500) and $\Upsilon=3.81$ (CLIC3000)~\cite{Group:2004sz}.
Notice how the improved CYC approximation considerably overestimates the effect 
of beamstrahlung at low $x$.
We then implemented the numerical solution in the CompHEP event generator~\cite{Pukhov:1999gg}. 
The resulting true muon energy distribution including both types of beam energy losses 
is shown in Fig.~\ref{fig:numsol}b. We see that beamstrahlung still degrades the kinematic 
endpoint, although not as much as the CYC and YC approximations would suggest.

\begin{acknowledgments}
AD is supported by the US DoE and the 
Michigan Center for Theoretical Physics.
The work of KK and KM is supported in part by 
a US DoE Outstanding Junior Investigator 
award under grant DE-FG02-97ER41209.
\end{acknowledgments}


\end{document}